# Wavelength-Multiplexed Nonlinear Computing with a Single-Layer Diffractive Optical Processor

Yongkang Cheng[1,2,3†], Che-Yung Shen[1,2,3†], Yuntian Wang[1,2,3], Shiqi Chen[1,2,3], Aydogan Ozcan[1,2,3*]

[1]Electrical and Computer Engineering Department, University of California, Los Angeles, CA, 90095, USA

[2]Bioengineering Department, University of California, Los Angeles, CA, 90095, USA

[3]California NanoSystems Institute (CNSI), University of California, Los Angeles, CA, 90095, USA

[†]These authors contributed equally to the work

[*]Correspondence to: ozcan@ucla.edu

## Abstract

Diffractive optical processors provide a promising platform for high-throughput, low-latency analog computing by exploiting engineered wave propagation to transform optical fields. However, implementing nonlinear mappings in optical hardware remains challenging. Here, we introduce a wavelength-multiplexed encoding-and-decoding (E+D) diffractive processor that exploits multiple illumination wavelengths to enhance the nonlinear function-approximation capability of a compact single-layer diffractive architecture, without cascading multiple diffractive layers or increasing the trainable degrees of freedom. Under sequential wavelength scanning, each wavelength–detector channel implements a distinct nonlinear function at the output of the diffractive processor. Numerical simulations using 100 wavelength channels and 10,000 spatial detector regions at the output of the diffractive processor demonstrate the implementation of $10^6$ distinct nonlinear functions through wavelength multiplexing. We also demonstrate a simultaneous multiwavelength illumination configuration in which the E+D diffractive processor performs wavelength-division multiplexing and nonlinear function approximation concurrently. This multiwavelength illumination configuration is experimentally validated by optically implementing 64 distinct nonlinear functions under simultaneous illumination at eight wavelengths from 495 to 635 nm. Through numerical simulations and visible-light experimental demonstrations, our results indicate that this wavelength-multiplexed nonlinear function approximation framework provides a scalable route toward compact, high-capacity diffractive processors for large-scale analog optical computing and optical information processing.

## 1. Introduction

The rapid growth of data-intensive computing and machine-learning applications has created an increasing demand for hardware platforms that can process, transmit and transform large volumes of information at higher speeds and with greater energy efficiency. Modern artificial intelligence workloads, in particular, rely on massively parallel operations and repeated high-dimensional transformations, while the movement of data between memory, processors and communication links often imposes substantial latency and energy overhead[1–3]. These challenges have motivated growing interest in computing architectures that can complement conventional digital electronics by accelerating specific operations at the physical layer[4–8]. Optical computing might offer an appealing route in this direction, as light propagation naturally supports high-dimensional parallel information processing, low-latency signal transformation and high-bandwidth transmission. Within this broader landscape, diffractive optical processors or diffractive deep neural networks have emerged as a powerful framework for task-specific analog computation[9–12]. These processors consist of spatially engineered passive diffractive surfaces that process optical fields through free-space propagation. When optimized in a data-driven manner for a prescribed task, such passive optical systems can implement a broad range of input–output transformations with high fidelity[12–16]. Beyond conventional single-task operations, optical degrees of freedom, including polarization and wavelength diversity, have also been exploited to multiplex a large set of transformations within a shared diffractive processor[14,15], while more recent work has demonstrated reconfigurable multi-task diffractive computing through illumination-phase multiplexing[17].

An important element of advanced information processing is nonlinearity. Nonlinear transformations enable computing systems to represent complex relationships and approximate mappings that cannot be captured by linear operations alone[18]. This capability is particularly critical in machine learning, where nonlinear operations enable function approximation, hierarchical feature extraction, classification and inference from high-dimensional inputs[19]. Translating such nonlinear functionality into optical hardware, however, remains difficult. In many practical photonic platforms, intrinsic optical nonlinearities are relatively weak, and appreciable nonlinear interactions often require optical intensities that are impractical for scalable free-space systems. Strategies based on nonlinear or active materials and field-enhancing device structures can reduce the required optical power, but often introduce trade-offs involving optical loss, activation threshold, response speed, integration complexity and scalability[20–23]. These challenges have motivated diffractive optical processors to realize nonlinear computational mappings using only linear optical materials, with nonlinearity emerging from input encoding and coherent wave propagation rather than from intrinsic material nonlinearities[24–31].

Despite these advances, scaling the multiplexing capacity of diffractive nonlinear function approximators remains a challenge. Because multiple nonlinear functions must share a finite optical aperture and limited spatial degrees of freedom, increasing the number of multiplexed function mappings can lead to reduced approximation accuracy and increased crosstalk[26]. Therefore, processing substantially more independent nonlinear mappings within a compact optical processor, without relying solely on enlarging the output aperture, remains an important challenge for scalable analog optical computing.

Here, we introduce a wavelength-multiplexed encoding-and-decoding (E+D) diffractive processor architecture that leverages illumination-wavelength diversity to significantly enhance the nonlinear function-approximation capacity of a compact diffractive optical processor. The proposed architecture is highly compact, integrating input-dependent wavefront encoding and trainable optical decoding within a single diffractive layer, implemented, for example, on a single spatial light modulator (SLM). By exploiting multiple illumination wavelengths within the same E+D architecture (either sequentially or simultaneously), this diffractive processor enhances its nonlinear function-approximation capacity without cascading multiple diffractive layers or increasing the number of trainable spatial degrees of freedom. The effective nonlinearity with respect to the input variable arises from the phase encoding and square-law intensity detection; the intervening optical propagation remains linear in the complex field.

Under sequential multiwavelength illumination, different wavelength channels are processed one at a time, and each wavelength–detector channel implements a distinct nonlinear function, enabling the functional capacity of the processor to scale up jointly with the number of illumination wavelengths and output detectors. Using 100 wavelength channels and 10,000 spatial detector regions at the output plane of an E+D diffractive processor, large-scale numerical simulations demonstrate the successful implementation of one million independently generated and assigned nonlinear functions drawn from the selected Fourier-harmonic family. We also demonstrate a simultaneous multiwavelength illumination configuration in which the E+D processor performs wavelength division and nonlinear function approximation concurrently. We further experimentally validate this simultaneous multiwavelength illumination configuration using 8 wavelengths in the visible spectrum, successfully implementing 64 distinct nonlinear functions at the output of our E+D processor. Through large-scale numerical simulations and visible-light experimental demonstrations, our results and analyses indicate that wavelength multiplexing significantly enhances the nonlinear function-approximation capacity of E+D diffractive optical processors. More broadly, this wavelength multiplexing strategy offers a scalable path toward compact, high-capacity diffractive systems for large-scale analog optical computing and other optical information-processing applications.

Finally, we emphasize that the Fourier harmonics employed by the encoder of our diffractive architecture are mathematical basis functions of the input, rather than propagating plane waves or transverse spatial-frequency components of the optical field. Accordingly, this architecture is fundamentally distinct from classical Fourier-optical or holographic processors: rather than implementing a prescribed operation on the spatial-frequency content of an optical field, it uses input-domain harmonics together with a trainable, wavelength-dependent diffractive decoder and square-law detection to implement a large set of independently prescribed nonlinear input–output mappings that are wavelength-multiplexed.

## 2. Results

### 2.1. Nonlinear Function Approximation using a Wavelength-Multiplexed E+D Diffractive Processor under Sequential Multiwavelength Illumination

**Figure 1** illustrates the overall architecture of the wavelength-multiplexed E+D diffractive nonlinear function approximator under sequential multiwavelength illumination. The input variable $a$ is encoded as a phase profile $\varphi_{\mathcal{E}}(a)$, which consists of an array of $N_d$ encoder regions $\{\mathcal{E}_k\}_{k=1}^{N_d}$ at the input encoder plane, using, for example, an SLM. Within each encoder region, the phase distribution is $\varphi_{\mathcal{E}_k}(p;a) = 2\pi pa,\ p = 0, \dots, N_p - 1$, where $p$ denotes the order of the Fourier harmonic components and $N_p$ denotes the number of harmonic components that are used to synthesize the target nonlinear functions[25,30,31]. The E+D diffractive processor is sequentially illuminated by uniform plane waves at $N_w$ discrete wavelengths, $\{\lambda_w\}_{w=1}^{N_w}$. Since different wavelength channels are applied sequentially, the encoder phase distribution is defined directly at the illumination wavelength $\lambda_w$ used in each sequential pass. For each illumination wavelength $\lambda_w$, the encoded optical field is modulated by an input-independent decoder phase profile $\mathcal{D}_w(\theta)$, which spans the entire input modulation/SLM surface; here, $\theta$ denotes the $N$ trainable parameters of the decoder phase profile. $\mathcal{D}_w(\theta)$ is fixed for each illumination wavelength, and it is superimposed on the input-dependent encoder phase profile at the SLM plane. Stated differently, $\mathcal{D}_w(\theta)$ acts as an optimizable input-independent phase bias, and it is defined with respect to a reference wavelength $\lambda_{ref}$, i.e., $\mathcal{D}_w(\theta) = \lambda_{ref}/\lambda_w \times \mathcal{D}_{ref}(\theta)$. Input encoding and optical decoding are therefore co-localized within a single diffractive layer, as the name "E+D" diffractive processor indicates. Therefore, the E+D architecture is distinguished by its compactness, integrating input encoding and target-specific decoding into a single input-diffractive layer.

After free-space propagation (see **Fig. 1**), the resulting optical intensity distribution is sampled by $N_d$ predefined detector regions. Each detector provides a scalar optical readout by integrating the output intensity over its spatial extent. For each wavelength-detector channel, the resulting signal is normalized to the range [0, 1]. Because the wavelength channels are illuminated and read sequentially, the same detector array is reused for each illumination wavelength, implementing a distinct set of nonlinear functions at each $\lambda_w$. Consequently, a single-layer E+D diffractive processor is trained to implement $N_f = N_w N_d$ distinct functions within the same architecture. We denote each nonlinear function as $f_{\lambda_w,k}$ where $k = 1, \dots, N_d$ refers to the output detectors. The trainable decoder set $\mathcal{D}_w(\theta)$ is jointly optimized across all wavelength-detector channels using a deep-learning-based optimization framework with a customized mean-squared error (MSE)-based loss function (see the **Materials and Methods** section), such that the normalized detector output $\hat{f}_{\lambda_w,k}(a)$ approximates its assigned target nonlinear function: $f_{\lambda_w,k}(a) \approx \hat{f}_{\lambda_w,k}(a)$. In this design, wavelength multiplexing expands the function space of the E+D diffractive processor from purely spatial readout to a joint wavelength-detector channel space, enabling a single diffractive layer to approximate a significantly larger number of nonlinear functions ($N_f = N_w N_d$) than a monochromatic diffractive processor, without increasing the number of diffractive layers or scaling up detectors.

Initially, we numerically demonstrated the large-scale nonlinear function-approximation capacity of the wavelength-multiplexed E+D processor by approximating one million independently assigned nonlinear functions using a single diffractive layer. The output plane consisted of a $100 \times 100$ detector array, corresponding to $N_d = 10^4$ spatial detector channels; together with $N_w = 100$ wavelength channels uniformly distributed from 450 to 650 nm, this provided a total of $N_f = N_w N_d = 10^6$ wavelength–detector function channels. Each target nonlinear function was randomly constructed from $N_p = 4$ complex-valued Fourier harmonic components, and the encoder contained the corresponding four harmonic orders such that its harmonic basis matched that of the target functions (refer to the **Materials and Methods** for details). The input variables ($a$) were sampled over a domain of interest $\mathcal{U}[-0.5,0.5)$ during training. A shared decoder set, i.e., $\mathcal{D}_w(\theta) = \lambda_{ref}/\lambda_w \times \mathcal{D}_{ref}(\theta)$, was jointly optimized across all wavelength–detector channels, enabling distinct wavelength-dependent optical transformations and wavelength-multiplexed nonlinear-function approximation within the E+D processor. In these analyses, the longest illumination wavelength was selected as the decoder reference wavelength, i.e., $\lambda_{\text{ref}} = 650$ nm.

**Figure 2** summarizes the optimization and fine-tuning of this million-function E+D diffractive processor design. To execute $N_f = 10^6$ distinct nonlinear functions, we first trained the optical decoder using a relatively large learning rate of $5 \times 10^{-3}$ for 5,000 gradient steps. The resulting decoder was subsequently fine-tuned for an additional 2,500 steps using a reduced learning rate of $1 \times 10^{-4}$. Throughout both stages, an adaptive function-weighted MSE loss was employed to balance function-approximation performance across the one million nonlinear function channels to avoid some functions having poor approximation accuracy (see **Materials and Methods** for details). Every 250 steps, the function-approximation performance was evaluated using a fixed input grid. The checkpoint achieving the lowest evaluation MSE was retained in each stage, and the optimized/frozen decoder obtained after fine-tuning is shown in **Fig. 2(a).**

After the completion of this two-stage optimization, the wavelength-resolved function approximation results revealed comparable approximation accuracy across the full 450–650 nm operating spectral range (**Fig. 2(b)**). The worst, median, and best MSE values remain at similar levels across the 100 distinct wavelength channels, with no wavelength exhibiting a systematic degradation in its function approximation accuracy. These results demonstrate that the shared E+D processor can maintain both high approximation accuracy and relatively uniform performance across the input wavelengths while supporting $N_f = 10^6$ independently assigned nonlinear functions.

The fine-tuning step in our design produced a substantial improvement across the entire distribution of function-approximation errors (see **Fig. 2(c)**). Compared with the distribution obtained after the initial training stage outlined above, the post-fine-tuning MSE distribution shifted markedly toward lower values, with significant improvements observed in the worst, median, and best-performing function channels. Representative examples corresponding to these locations in the two distributions are shown in **Fig. 2(d)**, where the detector outputs exhibit increasingly close agreement with their assigned target nonlinear functions after the fine-tuning step. This improvement, therefore, reflects a broad enhancement in approximation accuracy across the same function ensemble rather than an improvement limited to a small subset of high-error function channels.

To analyze how the wavelength-multiplexing capacity scales with the E+D architecture size, or how the approximation accuracy degrades with additional wavelength channels, next we systematically varied both the spatial architecture and the decoder parameters of the E+D processor (see **Fig. 3**). Consistent with the one-million-function configuration ($N_f = 10^6$) reported in **Fig. 2**, the detector width $d$, the edge-to-edge detector spacing $\Delta d$, and the axial propagation distance $z$ were held fixed across our analysis. The number of detector channels $N_d$ was varied among 100, 400, and 1600, while the number of trainable decoder parameters $N$ was increased proportionally so that the ratio $N/N_d$ remained constant. For each spatial architecture, the number of wavelength channels $N_w$ was progressively increased from 4 to 128, i.e., $N_f = N_w N_d$ progressively increased. To examine the influence of target-function harmonic complexity, the same scalability analysis was also performed for nonlinear functions randomly constructed from $N_p = 4$ and $N_p = 16$ complex-valued Fourier harmonic components.

Based on these analyses, **Figs. 3(a), (c)** reveal a notable scaling behavior of the wavelength-multiplexed E+D processor. Increasing $N_d$ directly increases the number of spatially multiplexed function channels, and when $N_d$ is increased together with the number of trainable decoder parameters $N$, the diffractive processor can support a larger $N_w$ at the same function approximation-error level. Thus, spatial scaling not only increases the number of functions supported through spatial multiplexing but also enhances the usable wavelength-multiplexing capacity. This favorable coupling is observed for both $N_p = 4$ and $N_p = 16$, suggesting that wavelength multiplexing is particularly well suited for scaling the function capacity of large-scale E+D diffractive processors.

**Figures 3(b),(d)** further reveal a complementary trade-off between spatial and wavelength multiplexing. At a fixed total number of nonlinear functions, $N_f = N_w N_d$, E+D diffractive processor configurations with larger $N_d$ and correspondingly smaller $N_w$ generally achieve lower function approximation errors. Increasing $N_d$, however, requires a proportional expansion of the spatial encoder and detector array regions together with the trainable decoder size. In contrast, for fixed $N_d$ and $N$, increasing $N_w$ provides an efficient way to increase the total number of approximated nonlinear functions ($N_f$) without enlarging the spatial input or output apertures.

### 2.2. Nonlinear Function Approximation with a Wavelength-Multiplexed E+D Diffractive Processor under Simultaneous Multiwavelength Illumination

We next investigated whether a wavelength-multiplexed E+D processor could perform wavelength division and nonlinear-function approximation under simultaneous multiwavelength illumination. As illustrated in **Fig. 4(a)**, in this case, multiple wavelength channels simultaneously illuminate the E+D plane, and each detector is assigned a target wavelength and a distinct nonlinear function. Unlike the sequential multiwavelength illumination case considered above, the detector response under simultaneous illumination contains contributions from all the illumination wavelength channels. The processor must therefore reproduce the assigned nonlinear functions while suppressing intensity contributions from non-target wavelengths per detector.

For our numerical simulations to validate this concept, we considered 8 illumination wavelengths from 495 to 635 nm with a 20-nm interval and an $8 \times 8$ array of $N_d = 64$ detector regions. The 64 detectors approximate 64 independently assigned nonlinear-function channels, with eight detectors assigned to each wavelength. To enable wavelength-division multiplexing across these eight wavelengths, each encoder region $\mathcal{E}_k$was assigned a fixed encoder reference wavelength $\lambda_k^{\text{ref}}$, selected from the eight illumination wavelengths, as shown in **Fig. 4(b)**. The phase distribution $\varphi_{\mathcal{E}_k}(p; a)$ of each encoder region was defined at its assigned $\lambda_k^{\text{ref}}$ and was scaled as $\varphi_{\mathcal{E}_k,w}(p; a) = \frac{\lambda_k^{\text{ref}}}{\lambda_w} \varphi_{\mathcal{E}_k}(p; a)$ for each illumination wavelength $\lambda_w$. Each detector $D_k$ was spatially paired with its corresponding encoder region and assigned the target wavelength $\lambda_k^{\text{target}} = \lambda_k^{\text{ref}}$, forming the wavelength–detector mapping shown in **Fig. 4(c)**. Following the same principle as in the multiwavelength sequential illumination shown in **Figs. 1-2**, the shared decoder reference wavelength was set to the longest illumination wavelength ($\lambda_{\text{ref}} = 635$ nm). For each illumination wavelength, the total phase modulation imparted by the E+D plane is given by the sum of the wavelength-dependent encoder phase and decoder phase distributions (see **Materials and Methods**).

The shared decoder $\mathcal{D}(\theta)$ was jointly optimized for multiplexed nonlinear function approximation and wavelength-division. One of the training loss terms minimized the MSE between the detector outputs and their assigned target functions, while an additional inter-wavelength crosstalk term suppressed non-target wavelength contributions within each detector region (see **Materials and Methods**). Therefore, in this simultaneous illumination configuration, wavelength division and nonlinear-function approximation tasks are learned jointly within the same E+D diffractive layer.

Our analyses confirm that the optimized E+D processor achieved low function approximation errors across the 64 wavelength–detector channels (**Fig. 4(d)**) while maintaining effective wavelength division among the assigned detector regions (**Fig. 4(e)**). Here, the inter-wavelength crosstalk at each detector was quantified as the maximum fraction of the detected intensity contributed by non-target wavelengths over all the evaluated input $a$ values; thus, **Fig. 4(e)** reports the worst-case spectral crosstalk for each detector. These numerical results demonstrate that a single shared E+D processor can, at the same time, perform wavelength division multiplexing and nonlinear-function approximation under simultaneous multiwavelength illumination.

### 2.3. Experimental Demonstration of Nonlinear Function Approximation using a Wavelength-Multiplexed E+D Diffractive Processor under Simultaneous Multiwavelength Illumination

Following the numerical demonstration of nonlinear-function approximation under simultaneous multiwavelength illumination, we experimentally implemented the corresponding wavelength-multiplexed E+D processor in a visible-light optical system. As illustrated in **Fig. 5(a)**, multiple illumination wavelengths were directed by a beam splitter (BS) onto a phase-only SLM implementing the combined E+D phase modulation. The modulated optical field was reflected from the SLM and propagated through free space onto the camera plane, where the resulting intensity distribution was recorded and sampled using the predefined detector regions.

In practice, the physical optical system inevitably deviates from the numerical forward model because of wavelength-dependent device responses, optical aberrations, alignment errors, illumination variations, detector nonuniformity, and other hardware-dependent imperfections. Consequently, directly displaying the numerically optimized phase patterns on the SLM results in degraded function-approximation performance relative to the numerical predictions (see **Fig. 6**). To compensate for these discrepancies, we employed in situ learning to further optimize the decoder using experimentally measured outputs[32–35]. Starting from the numerically optimized decoder phase profile, candidate phase patterns were evaluated on the physical system, and the measured nonlinear function approximation error was used to iteratively update the decoder policy. This measurement-driven, model-free optimization therefore directly accounts for the aggregate response of the physical system without requiring that all residual experimental imperfections be explicitly modeled/quantified (see **Materials and Methods** for details). The checkpoint corresponding to the minimum error during this in situ learning process was retained as the final experimental decoder, whose phase pattern is shown in **Fig. 5(c)**. As shown in **Fig. 5(d)**, the experimentally measured approximation error progressively decreased during in situ optimization, from an initial, zero-shot MSE of 0.0212 to a minimum of 0.0061 at epoch 195. The improvement was observed across all eight wavelength channels, with the per-wavelength MSE consistently reducing after in situ optimization (**Fig. 5(e)**). The measured responses of all 64 wavelength–detector channels likewise show substantially improved agreement with their corresponding target nonlinear functions after optimization (**Fig. 6**). These results experimentally demonstrate simultaneous multiwavelength nonlinear-function approximation using a single shared E+D processor and show that in situ learning effectively compensates for residual discrepancies between the numerical model and the physically deployed optical processor.

### 2.4. Quantitative Comparison of Sequential and Simultaneous Multiwavelength Illumination Configurations

The preceding experimental results demonstrate the feasibility of simultaneous multiwavelength nonlinear-function approximation using a shared E+D processor. However, the simultaneous multiwavelength diffractive processor configuration introduces additional constraints that are absent when wavelength channels are processed sequentially. In particular, in the simultaneous illumination case, the output detectors integrate contributions from multiple illumination wavelengths, and all wavelength channels must share the same decoder. To better understand these performance trade-offs, we numerically compared three E+D diffractive processor configurations with progressively reduced wavelength-sharing constraints: (i) the simultaneous multiwavelength configuration, corresponding to our experimental setup reported in the earlier section; (ii) an ideal wavelength-filtered configuration in which a bandpass filter at each detector transmits only its assigned wavelength; and (iii) a sequential illumination configuration (see **Materials and Methods** for details).

In these quantitative analyses, the simultaneous multiwavelength illumination configuration exhibited the highest approximation error across all eight wavelength channels, whereas the ideal wavelength-filtered configuration consistently achieved lower MSE, as shown in **Fig. 7**. The independently trained sequential illumination configuration provided the best overall performance.

This performance ordering remained consistent across the investigated illumination wavelength range, indicating that the readout condition and the degree of decoder sharing jointly influence the achievable accuracy of nonlinear function approximation.

The improvement from the simultaneous multiwavelength illumination configuration to the ideal wavelength-filtered configuration reflects the combined benefits of wavelength-selective readout and optimization under spectrally isolated/filtered detection conditions. Under the simultaneous multiwavelength illumination configuration, each output detector receives contributions from all illumination wavelengths, requiring the shared decoder to reproduce the target nonlinear functions while also suppressing contributions from non-target (undesired) wavelengths. In the ideal wavelength-filtered configuration, we assumed that these non-target contributions were filtered out by pre-assigned spectral bandpass filters positioned on the corresponding output detectors, thereby reducing the function-approximation error due to spectral crosstalk. The best MSE values were achieved with the sequential multiwavelength illumination configuration, in which each wavelength was assigned its own independently optimized decoder phase profile. Compared with the ideal wavelength-filtered simultaneous illumination configuration, this removes the additional constraint of sharing a common decoder profile across different wavelength channels.

Together, these comparisons shed light on how wavelength-integrated detection and shared-decoder optimization shape the achievable performance of nonlinear function approximation in the simultaneous multiwavelength configuration, while highlighting the trade-off between parallel multiwavelength processing and approximation fidelity.

## 3. Discussion and Conclusion

This work demonstrates wavelength multiplexing as an effective strategy for scaling the function capacity of diffractive nonlinear function approximators. Previous approaches have primarily relied on spatially separated detector channels, making the number of supported functions strongly dependent on the available output field of view and spatial degrees of freedom. Here, the E+D architecture extends nonlinear function approximation into a joint wavelength-detector channel space, where different wavelength–detector combinations can represent distinct nonlinear functions. Our large-scale numerical results demonstrate up to one million wavelength–detector function channels, while the visible-light experiment provides a proof-of-concept demonstration of the simultaneous multiwavelength illumination configuration across eight visible wavelength channels and 64 wavelength-assigned detector outputs, each corresponding to a distinct nonlinear function. Our presented demonstrations use a single scalar input and nonnegative target functions generated from a finite Fourier-harmonic family matched to the encoder basis. Extensions to multidimensional inputs, signed outputs, and target functions outside this finite basis remain to be investigated.

The practical implementations of our wavelength-multiplexed E+D diffractive processor design can be affected by optical aberrations, detector nonuniformity, alignment errors, system dispersion, and other discrepancies between the numerical model and the physical optical system. The model-free in situ optimization used in this work provides a complementary mechanism to compensate for model mismatches and hardware imperfections that are difficult to characterize explicitly.

The SLM-based visible-light E+D processor platform provides a flexible testbed for validating wavelength-multiplexed nonlinear function approximation. Further improvements could arise from optimizing the spatial architecture, propagation distance, wavelength-channel arrangement, and available degrees of freedom, as well as from more accurate characterization of wavelength-dependent system responses. Beyond the current SLM implementation, future systems could employ faster phase modulators, wavelength-addressable source arrays, integrated photonic devices, or fabricated diffractive structures to enable more compact, stable, and application-specific implementations[4,6,30,36–38].

More broadly, the wavelength-multiplexing strategy demonstrated here can be extended by incorporating additional physical degrees of freedom, such as polarization, spatial mode, propagation angle, and/or time multiplexing[14,17,39–44]. Jointly exploiting multiple optical dimensions could further expand the accessible function space without relying solely on increased spatial footprint. Such diverse multiplexing approaches provide a scalable path toward compact, high-capacity analog optical processors for computational imaging, sensing, and machine-learning-related information processing.

## 4. Materials and Methods

### Training Loss Functions and Performance Metrics

Different training objectives were used for the one-million-function numerical demonstration and the configurations presented in **Fig. 4** and **Fig. 7**. For the one-million-function model, an adaptive function-weighted MSE loss was employed to dynamically increase the loss contributions of the function channels with above-average fitting errors; this approach helped us avoid having poor approximation performance in outlier nonlinear functions. For the simultaneous multiwavelength illumination configuration, the function-approximation loss was evaluated without any adaptive weighting, while an additional inter-wavelength crosstalk loss was introduced for the simultaneous multiwavelength illumination configuration.

For the one-million-function numerical model shown in **Fig. 2**, each wavelength–detector pair $(\lambda_w, D_k)$ defines an independent target nonlinear function to be approximated. For a mini-batch containing $B$ input arguments $\{a_b\}_{b=1}^{B}$ , the loss of the target function channel associated with wavelength $\lambda_w$ and detector $D_k$ was calculated using:

$$\mathcal{L}_{w,k}^{\mathrm{MSE}} = \frac{1}{B}\sum_{b=1}^{B}\left[\hat{f}_{\lambda_w,k}(a_b) - f_{\lambda_w,k}(a_b)\right]^2,$$

where $f_{\lambda_w,k}(a_b)$ denotes the target nonlinear-function value and $\hat{f}_{\lambda_w,k}(a_b)$ denotes the corresponding diffractive output that is min–max normalized based on the intensity outputs $I_{\lambda_w,k}(a)$, i.e.,

$$\hat{f}_{\lambda_w,k}(a) = \frac{I_{\lambda_w,k}(a) - I_{\lambda_w,k}^{\min}}{I_{\lambda_w,k}^{\max} - I_{\lambda_w,k}^{\min} + \epsilon},$$

where

$$I_{\lambda_w,k}^{\min} = \min_a I_{\lambda_w,k}(a), \quad I_{\lambda_w,k}^{\max} = \max_a I_{\lambda_w,k}(a).$$

and a small positive constant $\epsilon = 10^{-9}$ was added to the denominator for numerical stability.

Because a large number of independently assigned nonlinear functions are optimized simultaneously, an adaptive weight $\alpha_{w,k}$ was assigned to every wavelength–detector function channel. All these adaptive weights were initialized to 1, such that the initial objective is equivalent to the unweighted MSE. The adaptive-weighted function-approximation loss was defined as

$$\mathcal{L}_{\text{adaptive}} = \frac{1}{N_w N_d} \sum_{w=1}^{N_w} \sum_{k=1}^{N_d} \alpha_{w,k} \, \mathcal{L}_{w,k}^{\text{MSE}}.$$

At each training iteration $t$, the reference loss $\mathcal{L}_{\text{ref}}$ was calculated as the unweighted MSE over all $N_w N_d$ function channels,

$$\mathcal{L}_{\text{ref}}^{(t)} = \frac{1}{N_w N_d} \sum_{w=1}^{N_w} \sum_{k=1}^{N_d} \mathcal{L}_{w,k}^{\text{MSE},(t)}.$$

The adaptive weight of each function channel was then updated according to

$$\alpha_{w,k}^{(t+1)} = \max\left[\alpha_{w,k}^{(t)} + \eta_\alpha \left(\mathcal{L}_{w,k}^{\text{MSE},(t)} - \mathcal{L}_{\text{ref}}^{(t)}\right), 0\right],$$

where $\eta_\alpha = 0.1$ denotes the adaptive-weight update rate. Function channels whose approximation errors exceed the global mean loss, therefore, receive increased weights in the total loss function, whereas those with below-average errors receive reduced weights. The non-negativity constraint prevents individual function weights from becoming negative. The updated weights $\alpha_{w,k}^{(t+1)}$ are used in the subsequent optimization iteration.

Because the update is additive and proportional to the difference between the individual function loss $\mathcal{L}_{w,k}^{\text{MSE}}$ and the global mean loss $\mathcal{L}_{\text{ref}}$, the magnitude of the weight adjustment naturally decreases as the fitting errors converge. When wavelength channels were processed in separate chunks for memory-efficient gradient accumulation, the reference loss $\mathcal{L}_{\text{ref}}$ and adaptive-weight update were evaluated only after all wavelengths in the current optimization step had been processed, ensuring that $\mathcal{L}_{\text{ref}}$ was calculated over the complete set of target function channels.

The adaptive weights $\alpha_{w,k}$ were initialized only at the beginning of the first training stage. Both the optimized decoder parameters and the adaptive-weight field were retained in the training checkpoint and carried into the subsequent low-learning-rate fine-tuning stage. The same adaptive-weighted objective and weight-update rule were therefore continued during fine-tuning without reinitializing $\alpha_{w,k}$.

For all configurations presented in **Fig. 4** and **Fig. 7**, the adaptive function weights $\alpha_{w,k}$ introduced for the one-million-function numerical model were not used. Instead, all detector channels contributed uniformly to the function-approximation loss. Each detector $D_k$ was assigned an independently prescribed target function $f_k(a)$, and its normalized optical output $\hat{f}_k(a)$ was obtained from min–max-normalization. For a mini-batch containing $B$ input arguments $\{a_b\}_{b=1}^{B}$, the function-approximation loss was calculated as

$$\mathcal{L}_{\text{approx}} = \frac{1}{N_d B} \sum_{k=1}^{N_d} \sum_{b=1}^{B} \left[\hat{f}_k(a_b) - f_k(a_b)\right]^2,$$

where $N_d$ is the total number of detectors, $f_k(a_b)$ is the target nonlinear-function value assigned to detector $D_k$, and $\hat{f}_k(a_b)$ is the corresponding normalized optical output.

For the simultaneous multiwavelength illumination configuration, an additional inter-wavelength crosstalk loss was introduced to suppress the contributions of non-target (i.e., undesired) wavelengths within each detector region. As defined above, detector $D_k$ was assigned a target wavelength $\lambda_k^{\text{target}}$, while $I_{\lambda_w,k}(a_b)$ denotes the integrated intensity contributed by wavelength $\lambda_w$ to detector $D_k$ at input $a_b$. For each detector and input argument, the fraction of the detected intensity contributed by the assigned target wavelength can be written as:

$$r_{\lambda_k^{\text{target}},k}(a_b) = \frac{I_{\lambda_k^{\text{target}},k}(a_b)}{\sum_{w=1}^{N_w} I_{\lambda_w,k}\,(a_b) + \epsilon},$$

where $N_w$ is the number of simultaneously active wavelength channels and $\epsilon = 10^{-9}$ is a positive constant introduced for numerical stability.

Rather than averaging this target-wavelength fraction over the mini-batch, the minimum value across the $B$ input arguments was used for each detector, corresponding to the input condition with the poorest wavelength separation. The inter-wavelength crosstalk loss was therefore defined as

$$\mathcal{L}_{\text{cross}} = \frac{1}{N_d} \sum_{k=1}^{N_d} \left[1 - \min_{b=1,\ldots,B} r_{\lambda_k^{\text{target}},k}(a_b)\right]$$

This worst-case formulation of spectral crosstalk penalizes the input condition exhibiting the lowest target-wavelength fraction at each detector, preventing poor wavelength separation at individual operating points from being obscured by averaging over the mini-batch. The same worst-case spectral crosstalk definition was used for the metric reported in **Fig. 4(e)**, where the maximum non-target-wavelength intensity fraction was evaluated for each detector over the sampled input $a$ values. The inter-wavelength crosstalk loss term evaluates wavelength mixing only within the predefined detector regions; optical power outside these regions is not included in this loss.

The total training objective for the simultaneous illumination wavelength-multiplexed model was

$$\mathcal{L}_{\text{total}} = \mathcal{L}_{\text{approx}} + \beta_{\text{cross}}\mathcal{L}_{\text{cross}},$$

where the relative weight of the inter-wavelength crosstalk loss term was set to

$$\beta_{\text{cross}} = 0.01.$$

Thus, $\mathcal{L}_{\text{approx}}$ optimizes the nonlinear-function approximation accuracy obtained from the wavelength-integrated detector signal, whereas $\mathcal{L}_{\text{cross}}$ uses the wavelength-resolved intensities available in the numerical forward model to promote wavelength purity at each detector. The independently trained single-wavelength sequential illumination configuration did not require the inter-wavelength crosstalk term ($\mathcal{L}_{\text{cross}}$) because only one wavelength was present at a time; in that case, optimization was performed using $\mathcal{L}_{\text{approx}}$ alone ($\mathcal{L}_{\text{total}} = \mathcal{L}_{\text{approx}}$). The ideal wavelength-filtered illumination configuration used the same wavelength-multiplexed optical architecture described above but was independently optimized under the wavelength-filtered readout condition, with only the target-wavelength contribution retained at each detector; therefore $\mathcal{L}_{\text{cross}}$ was not needed.

**Supplementary Information:**

- Optical Forward Model of a Wavelength-Multiplexed E+D Nonlinear Function Approximator
- Details of the Visible-Light Experimental Setup.
- Data Preparation and Other Numerical Implementation Details

## Figures

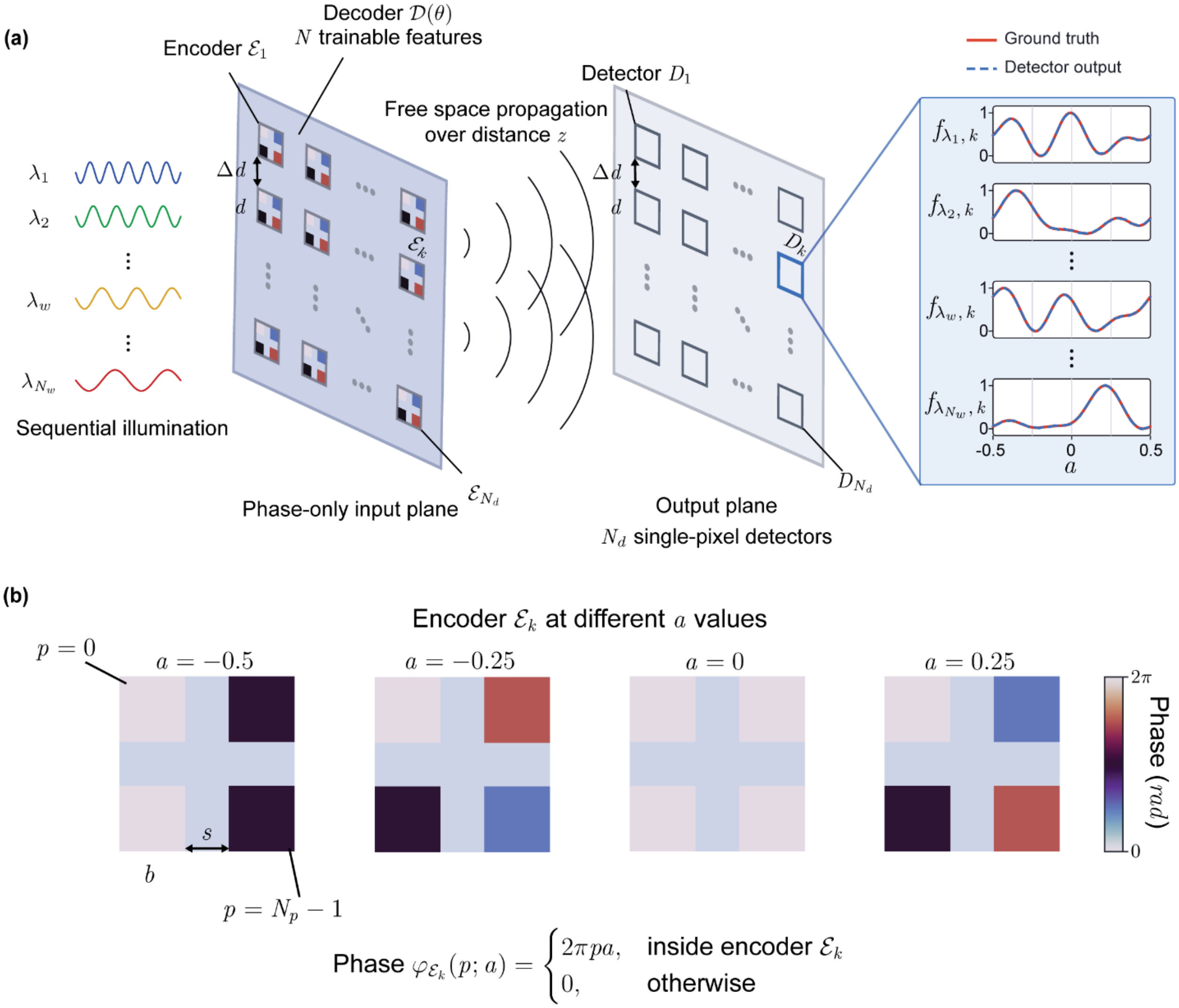

**Figure 1. Wavelength-multiplexed E+D diffractive processor for nonlinear function approximation. (a)** Schematic of the E+D processor. A set of spatially coherent plane waves at discrete wavelengths $\{\lambda_w\}_{w=1}^{N_w}$ sequentially illuminates a phase-only input plane containing $N_d$ encoder regions $\{\mathcal{E}_k\}_{k=1}^{N_d}$ and a shared trainable decoder phase profile $\mathcal{D}(\theta)$ with $N$ trainable features. Each encoder region has a width $d$, with an edge-to-edge spacing $\Delta d$ between adjacent regions. After free-space propagation over an axial distance $z$, the resulting optical intensity is measured by $N_d$ single-pixel detectors $\{D_k\}_{k=1}^{N_d}$, whose dimensions and spatial arrangement match those of the encoder regions. Under illumination at a wavelength of $\lambda_w$, the signal of the output detector $D_k$ provides the optical approximation of a predetermined nonlinear function $f_{\lambda_w,k}(a)$; this wavelength multiplexing scheme is used to approximate $N_f = N_w N_d$ distinct nonlinear functions at the output of the E+D diffractive processor. **(b)** Input encoding within an individual encoder region $\mathcal{E}_k$. The scalar input $a$ is mapped to $N_p$ distinct Fourier harmonic components. Representative phase distributions are shown here for different values of $a$. Each harmonic

component occupies a square sub-block of width $b$, with adjacent sub-blocks separated by a spacing $s$.

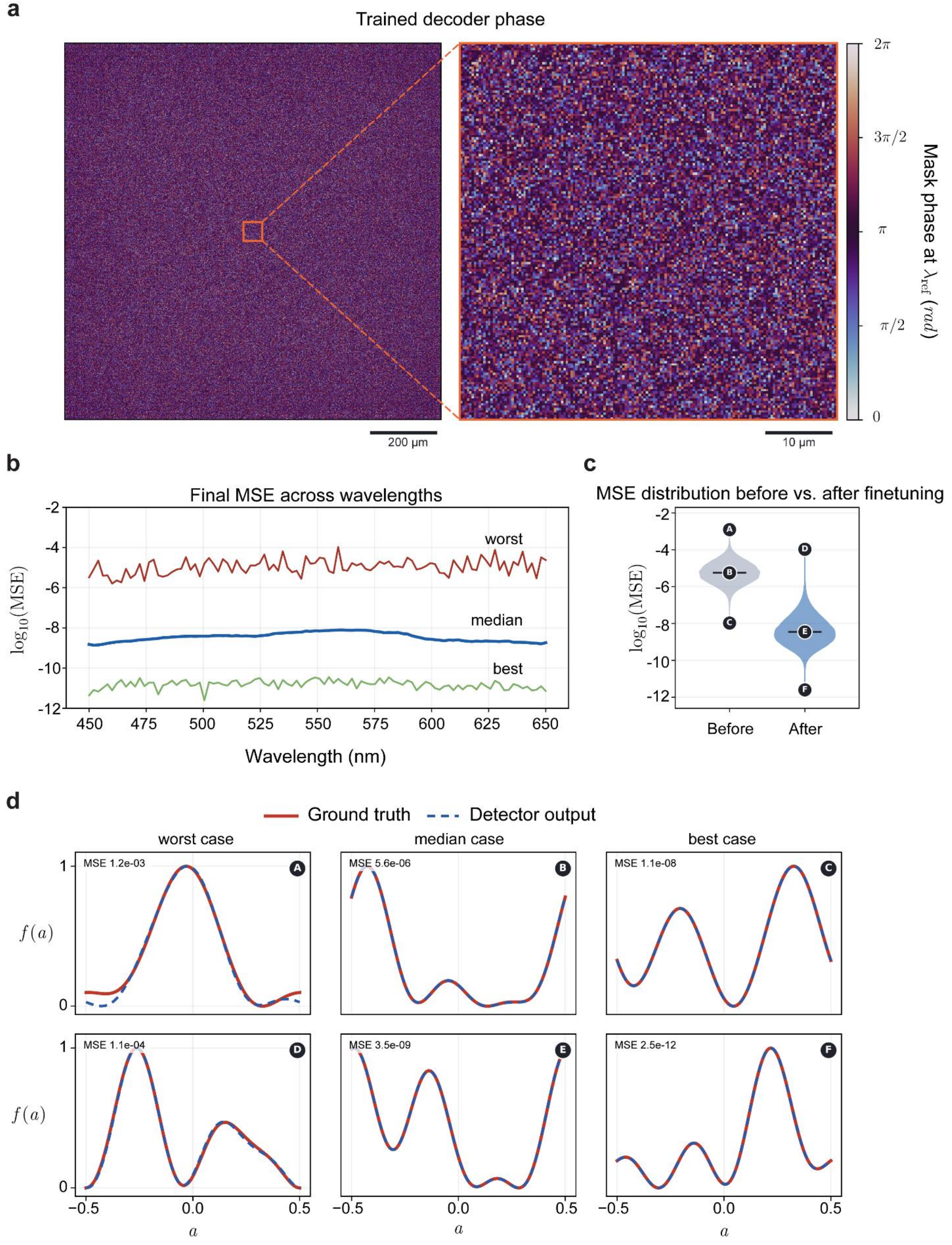

**Figure 2. Numerical demonstration of the approximation of one million distinct nonlinear functions using a wavelength-multiplexed E+D processor; $N_f = 10^6$.** The E+D processor was

jointly optimized over $N_w = 100$ wavelength channels and a $100 \times 100$ detector array, corresponding to $N_f = 10^6$ nonlinear function channels. Training was performed in two stages, with an initial learning rate of $5 \times 10^{-3}$ followed by fine-tuning at $1 \times 10^{-4}$. **(a)** Phase profiles of the optimized decoder. **(b)** The worst, median, and best MSE values across different wavelengths. **(c)** MSE distributions across all $10^6$ function channels before and after fine-tuning. Points (A–C) indicate the worst, median, and best cases before fine-tuning, respectively, while (D–F) indicate the worst, median, and best cases after fine-tuning. **(d)** Representative nonlinear function approximations associated with A–F in **(c)**, showing the target functions and corresponding detector outputs before (A–C) and after (D–F) fine-tuning.

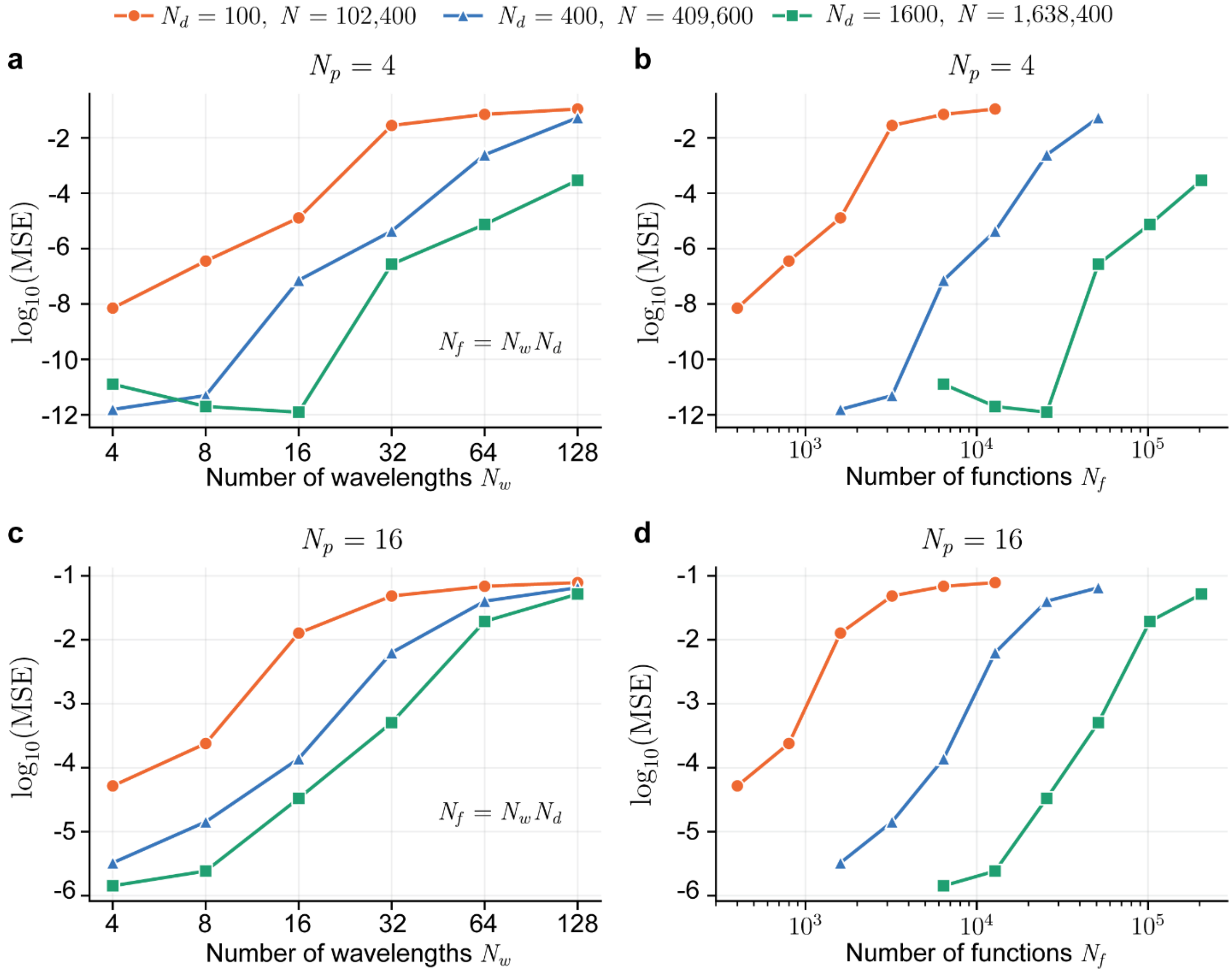

**Figure 3. Scalability analysis of the wavelength-multiplexed E+D processor.** The number of wavelength channels $N_w$ was increased from 4 to 128 for three different detector-array sizes, corresponding to $N_d = 100$, 400, and 1600 detectors and $N = 102{,}400$, 409,600, and 1,638,400 trainable decoder parameters, respectively. Function-approximation performance is reported as $\log_{10}(\mathrm{MSE})$. **(a), (c)** Approximation error as a function of the number of wavelength channels $N_w$ for target nonlinear functions generated as the squared magnitude of a sum of $N_p = 4$ and $N_p = 16$ complex-valued Fourier basis terms, respectively. **(b), (d)** Approximation error as a function of the total number of independently assigned nonlinear functions, $N_f = N_w N_d$, for target functions constructed with $N_p = 4$ and $N_p = 16$, respectively.

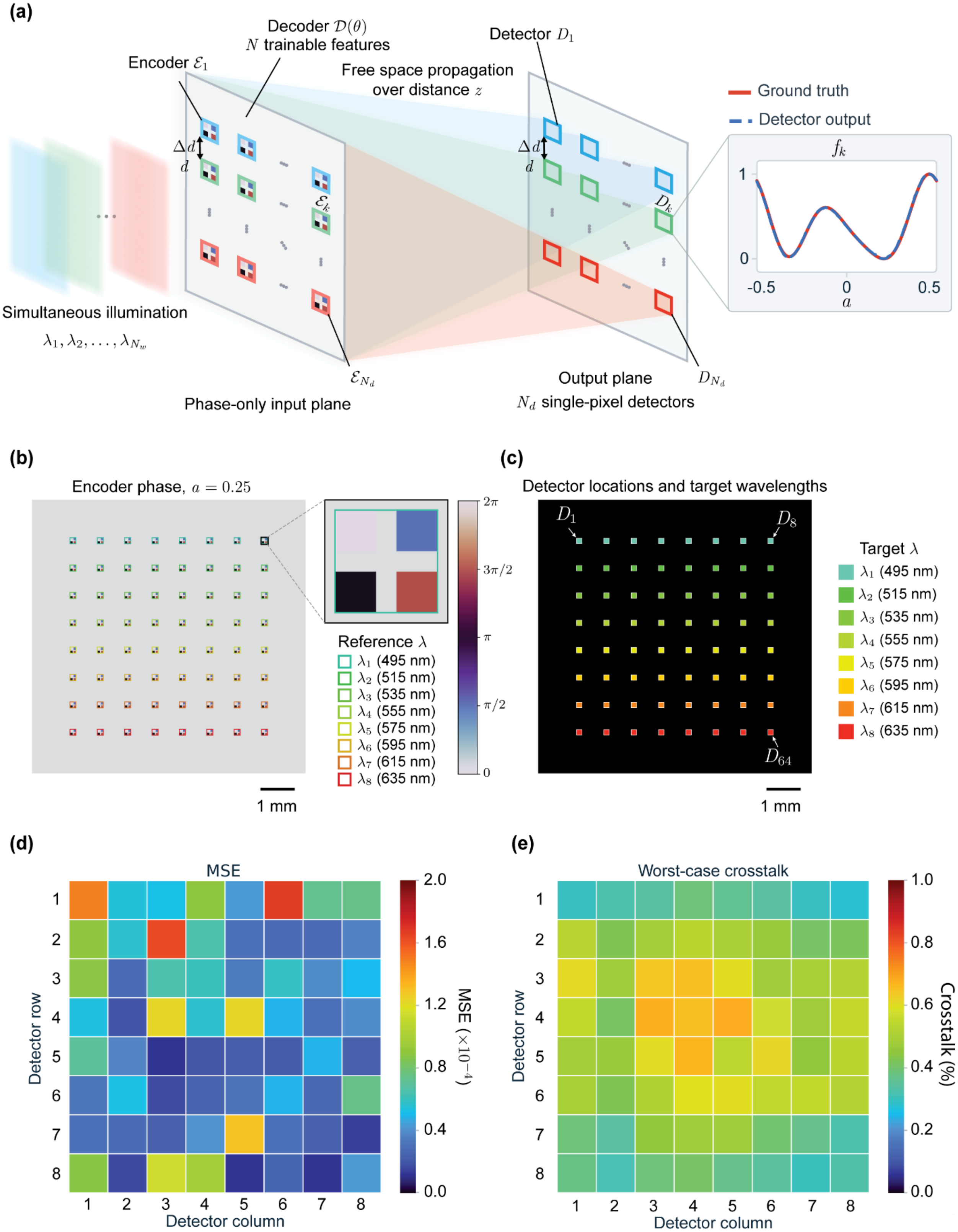

**Figure 4. Nonlinear-function approximation under simultaneous multiwavelength illumination using a wavelength-multiplexed E+D processor. (a)** Schematic of the simultaneous illumination-based multiwavelength E+D architecture, in which wavelength-

assigned encoder regions and a shared decoder jointly perform wavelength division and nonlinear-function approximation. **(b)** Encoder phase pattern for the simulated configuration with $N_d = 64$, with eight illumination wavelengths from 495 to 635 nm that are simultaneously active, and $N_p = 4$. **(c)** Corresponding detector arrangement. Each detector is assigned one target wavelength and one nonlinear function to be approximated. **(d)** Per-detector function-approximation MSE. **(e)** Per-detector worst-case spectral crosstalk (%).

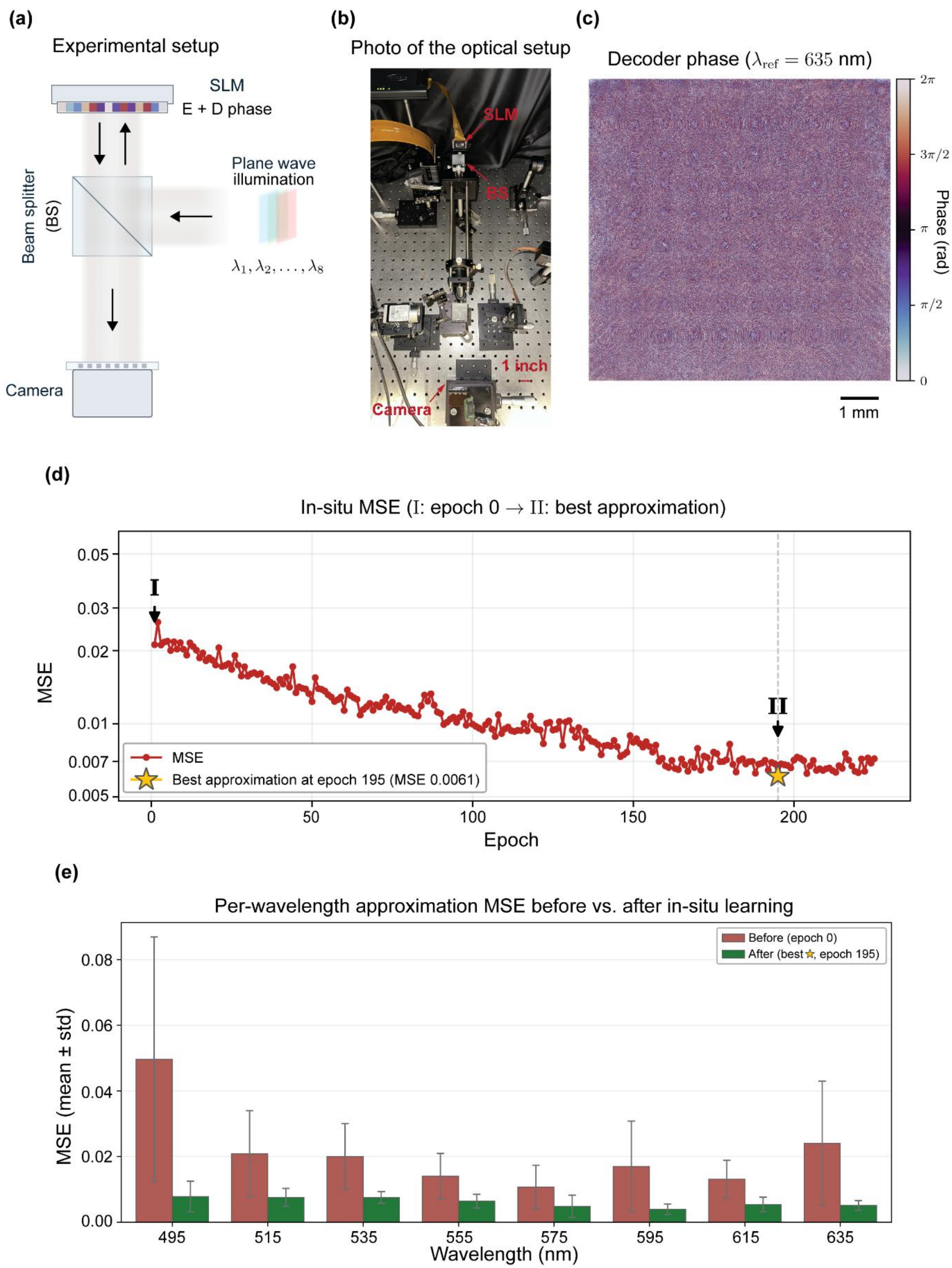

**Figure 5. Experimental demonstration of nonlinear function approximation using a wavelength-multiplexed E+D processor.** **(a)** Schematic of the experimental setup for

simultaneous multiwavelength illumination. (**b)** Photo of experiment setup. **(c)** Decoder phase profile at the reference wavelength $\lambda_{\mathrm{ref}} = 635$ nm after in situ learning. **(d)** Experimentally measured function-approximation MSE values during the PPO-based in situ optimization. Points I and II denote epochs 0 and 195, respectively; the latter (epoch 195) achieved the lowest MSE during in situ learning. **(e)** Per-wavelength function-approximation MSE before and after in situ optimization.

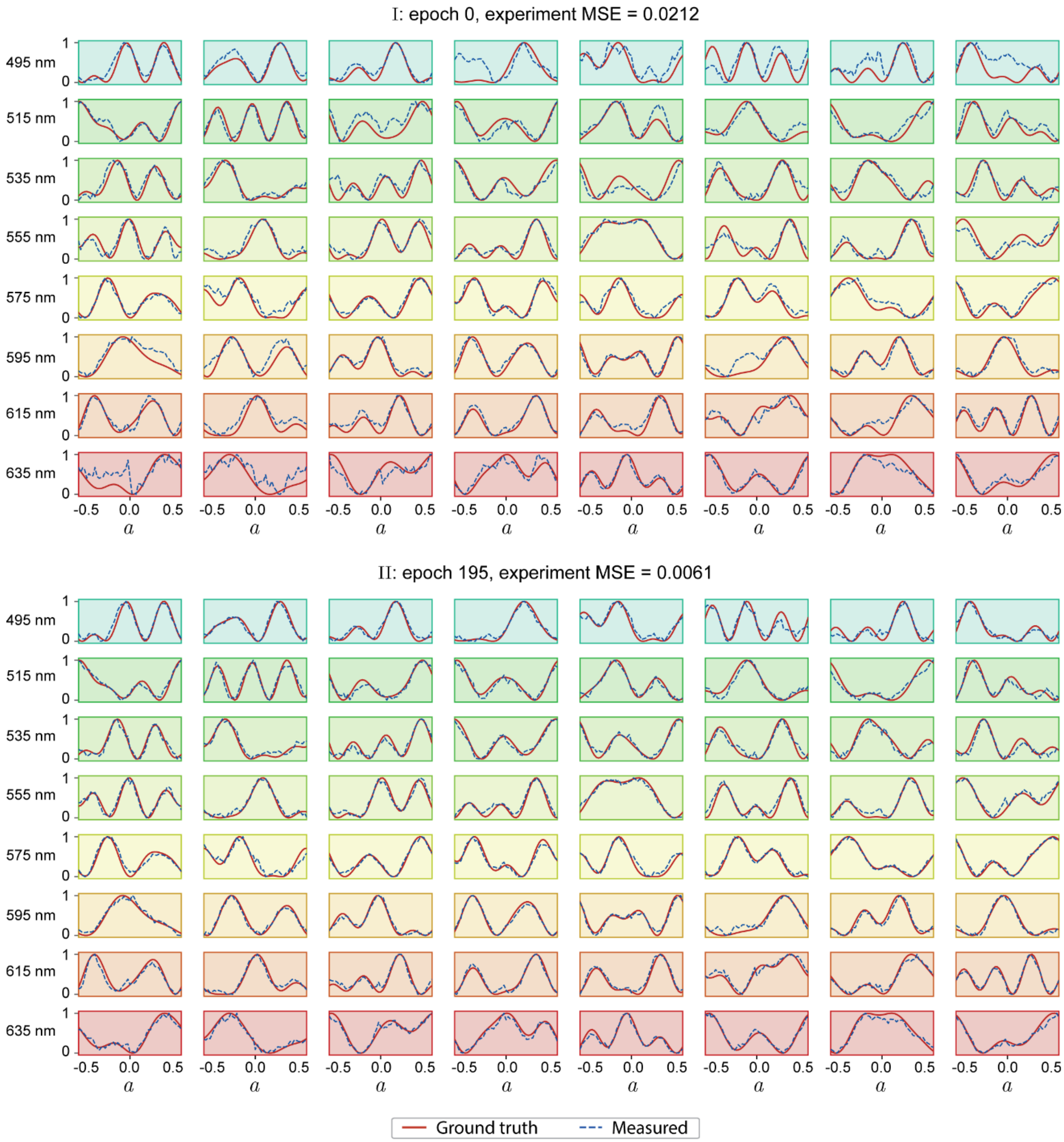

**Figure 6. Performance analysis of the experimental nonlinear-function approximation before and after PPO-based in situ optimization.** Normalized experimental outputs from the 64 detector channels, along with their corresponding target functions, over the target input range $a \in [-0.5, 0.5]$. Each subplot corresponds to a distinct detector and its assigned nonlinear function, and the spatial arrangement of the subplots follows the physical $8 \times 8$ detector layout at the output plane of the experimental setup. Detector rows are assigned to the eight illumination-wavelength channels spanning 495 to 635 nm, as indicated on the left. Red solid curves represent the target functions (ground truth), and blue dashed curves represent the experimentally measured and

normalized optical outputs under simultaneous eight-wavelength illumination. The upper and lower panels show experimental responses before and after the PPO-based in situ optimization, respectively.

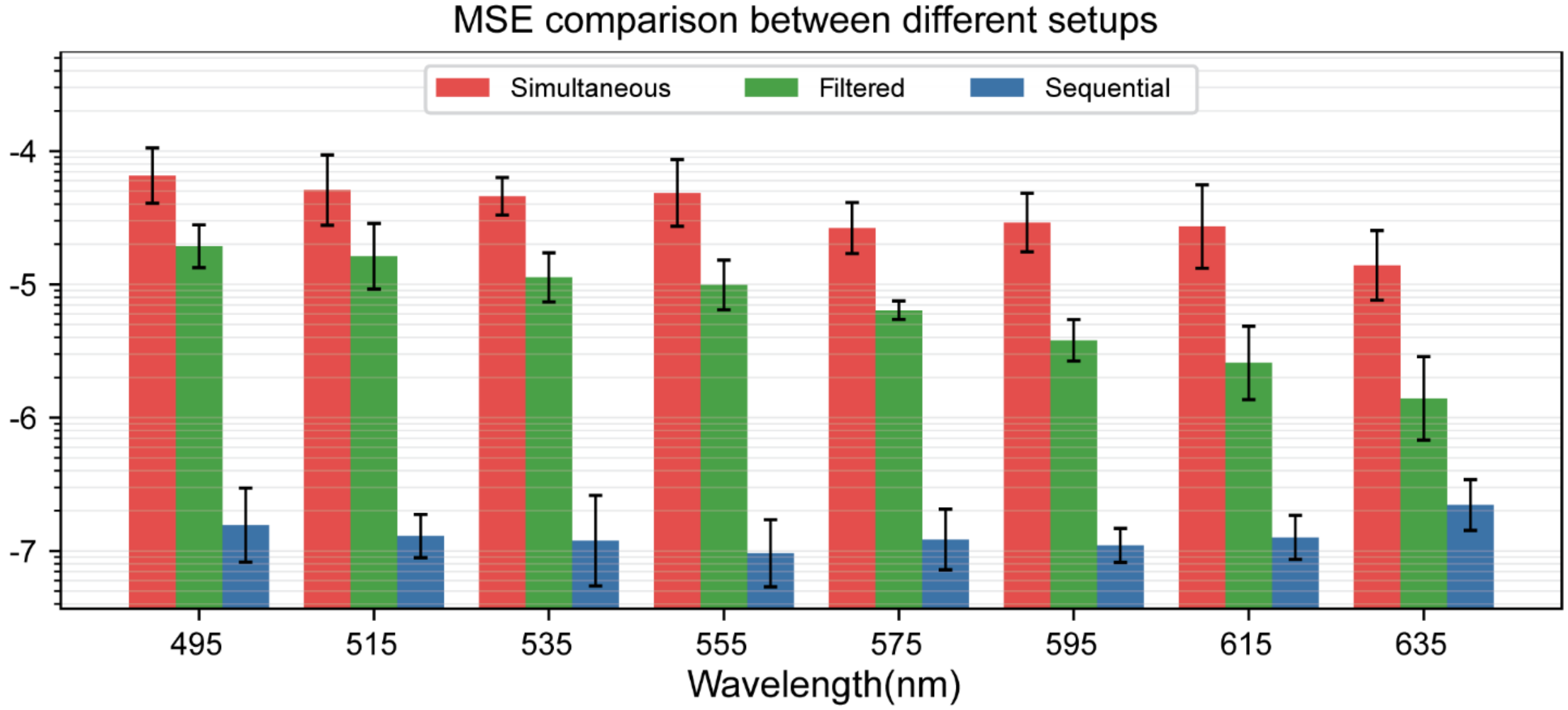

**Figure 7. Comparison of diffractive nonlinear function approximation errors across simultaneous, filtered, and sequential wavelength-readout configurations.** Function-approximation performance of the wavelength-multiplexed E+D processor is reported as $\log_{10}(\mathrm{MSE})$ for eight wavelength channels from 495 to 635 nm for the simultaneous multiwavelength illumination configuration, ideal wavelength-filtered configuration, and independently trained single-wavelength sequential illumination configuration. Bars show the mean $\log_{10}(\mathrm{MSE})$ of the function channels assigned to each wavelength, and error bars indicate the corresponding standard deviation in each case.